\documentclass[a4paper,12pt]{article}
\usepackage{graphicx} 
\usepackage{cite}
\usepackage{amsmath,amsfonts,amssymb}
\usepackage[small,bf,hang]{caption}
\usepackage{slashed}
\usepackage{color}
\usepackage{breqn}
\usepackage{physics}
\usepackage{extarrows}
\usepackage{graphicx}
\usepackage{dcolumn}
\usepackage{bm}
\usepackage{hyperref}
\usepackage{ulem}
\usepackage{geometry}
\newcommand{\be}{\begin{equation}}
\newcommand{\ee}{\end{equation}}
\newcommand{\ba}{\begin{eqnarray}}
\newcommand{\ea}{\end{eqnarray}}

\begin{document}

\begin{titlepage}

\vspace*{1.0cm}

\begin{center}
{\textbf{\huge Holographic partition function of democratic M-theory
}}
\end{center}
\vspace{1.0cm}

\centerline{
\textsc{\large J. A.  Rosabal}
\footnote{jarosabal80@gmail.com}
}

\vspace{0.6cm}

\begin{center}
{\it Departamento de Electromagnetismo y Electrónica, Universidad de Murcia,
Campus de Espinardo, 30100 Murcia, Spain.}\\
\end{center}

\vspace*{1cm}

\begin{abstract}

We study the partition function associated with the democratic formulation of M-theory, focusing on its global definition and quantum properties. Using a path-integral representation that makes manifest the underlying cohomological structure, we analyze the coupled system of M-theory form fields $(A_3 + A_6)$, and the background fields $(C_4 + C_7)$, as well as their associated global transformations.
We show that the resulting description is naturally captured by a Heisenberg-type group reflecting the presence of a quadratic coupling between electric and magnetic degrees of freedom. This framework provides a transparent characterization of the global structure of the theory, clarifies the role of higher-form global symmetry, and allows for a consistent definition of the partition function in terms of higher-dimensional auxiliary manifolds.

\end{abstract}

\thispagestyle{empty}
\end{titlepage}

\setcounter{footnote}{0}

\tableofcontents

\newpage

\section{Introduction}

The current status of the description of chiral form fields in
d dimensions, and of democratic non-chiral forms, is split into two main approaches. The first approach implements a formal and purely quantum treatment of chiral form fields \cite{Witten:1991mm,Witten:1996hc,Belov:2006jd}. Partition functions in
d dimensions are expressed as path integrals on a 
(d+1)-dimensional manifold of a topological Chern–Simons theory. Although this framework admits an elegant holographic interpretation \cite{Belov:2006jd}, the additional dimension has no associated physical reality; rather, these partition functions should be regarded as path  integral representations of certain functionals.

Before proceeding, we emphasize that the appearance of higher-dimensional
manifolds in our construction should be understood purely as a technical
device.
The twelve- and thirteen-dimensional formulations employed in this work
provide convenient path-integral representations of the same quantum
functional, making manifest its global and cohomological structure, rather
than describing theories with additional physical degrees of freedom.
Different bulk realizations should therefore be regarded as equivalent
representations of the same boundary partition function, analogous to
different trivializations of a line bundle associated with a quantum
theory.

The second approach adopts a less formal and more classical perspective \cite{Maldacena:2001ss,Evnin:2023ypu}. Chiral form fields in 
d
dimensions, as well as democratic non-chiral forms, are described as edge modes of topological Chern–Simons or BF-like theories in one higher dimension. This description assigns physical significance to the bulk manifold, which makes it more difficult to provide a coherent physical interpretation without invoking the higher-dimensional space.

In this context, a reduction procedure has been developed \cite{Arvanitakis:2022bnr,Evnin:2023ypu} to obtain chiral and democratic actions directly in 
d
dimensions. While this procedure enhances the physical role of the 
(d+1)-dimensional manifold, its advantage is that it connects with established democratic descriptions in
d
dimensions \cite{Avetisyan:2022zza,Avetisyan:2021heg}. Nevertheless, it remains unclear how to implement this reduction consistently within an effective quantum framework.

These two approaches differ substantially, yet they complement each other in meaningful ways. Insights from the second approach can sometimes provide heuristic guidance for improving our understanding of the first. Taking more seriously the analogy between the two, one immediately identifies a gap in the first approach: it does not incorporate non-chiral fields or democratic actions. Although extending it to democratic actions involving higher-form fields with ordinary Abelian symmetry appears straightforward, the extension to actions with Chern–Simons terms, such as the M-theory action, and higher-group symmetries \cite{Benini:2018reh,DelZotto:2022joo} is far from trivial.

The aim of this work is to fill this gap and present a formal and purely quantum treatment of non-chiral form fields and democratic actions with Chern–Simons terms and higher-group symmetry, in a manner similar to that developed in \cite{Witten:1996hc} for the chiral compact scalar in 2d and the self-dual 3-form on the M5-brane. We have chosen M-theory as an example, but the procedure presented here can be applied to any theory of similar structure, such as democratic Maxwell–Chern–Simons theory in 5d,  type IIB $SL(2)$ democratic supergravity or democratic axion electrodynamics.

Throughout this work, several obstacles arise. The first one is the formulation of the democratic M-theory action itself. Upon reviewing the literature, we find several formulations of democratic actions. Some of them are not suitable for performing quantum mechanical calculations. The price for not having to manually impose the duality relation between electric and magnetic field strengths classically is, in some cases, too high. Non-polynomial or gauge-fixed actions, or the introduction of new degrees of freedom \cite{Pasti:1996vs,Pasti:1997gx,Bandos:1997gd,Buratti:2019guq,Sen:2019qit,Avetisyan:2022zza,Mkrtchyan:2019opf,Evnin:2022kqn,Mkrtchyan:2022xrm} are some of the features we aim to avoid in any quantum description, particularly in M-theory.

To avoid these unwanted features in some formulations, we adhere to the most conventional one \cite{Bergshoeff:2001pv} where electric and magnetic degrees of freedom appear on equal footing. After varying the action and manually imposing the duality relation, the equations of motion (e.o.m.) of the original non-democratic action are obtained. It may seem that what has been stated in the previous sentence would also prevent us from performing meaningful quantum mechanics. However, it turns out that this is not the case. A similar assumption is the starting point in  \cite{Witten:1996hc}. Although, since we are approaching this problem from a purely quantum mechanical point of view as in \cite{Witten:1996hc}, we do not have to care about varying and manually imposing  the duality constraint. In fact it is better to think about these actions, or classical pseudo-actions \cite{Bergshoeff:2001pv}, as purely quantum actions with no classical counterpart.  With the goal of establishing a comparison between the calculation of the partition function of the chiral scalar in 2d and our work, we briefly review this calculation after overcoming this obstacle.

We have not overcome this obstacle yet. If we stick to this characterization of what a "democratic theory" is, we encounter at least two different actions in the literature. The first one \cite{Heidenreich:2020pkc} (see eq. 5.59) does not contain a Chern-Simons term, but the coefficients in front of the kinetic terms are different:
\be
S_{\text{DEM}}=\int_{M_{11}}\frac{1}{3} F_4\wedge\star  F_4+\frac{1}{6} F_7\wedge\star F_7\ , \quad F_7=-\text{i}\star F_4.
\ee
The second one cannot be found in the literature in the way we will present it, but there are some hints in \cite{Fernandez-Melgarejo:2023kwk} (see eq. 5.13) suggesting that a democratic action for M-theory could also be written as:
\be
S_{\text{DEM}}=\int_{M_{11}}\frac{1}{4} F_4\wedge\star  F_4+\frac{1}{4} F_7\wedge\star F_7-\frac{1}{6}\text{i g}A_3\wedge  F_4\wedge  F_4\ , \quad F_7=-\text{i}\star F_4.
\ee
This formulation does include a Chern-Simons term, and the coefficients in front of the kinetic terms are equal.

Quantum calculations would favor the first action, without the Chern-Simons term. However, the formalism we develop here works perfectly in the presence of such terms. For this reason, we work with a family of democratic M-theory actions, parametrized by two parameters $\alpha$, and $\beta$,
\be
S_{\text{DEM}}=\int_{M_{11}}\alpha F_4\wedge\star  F_4+\beta F_7\wedge\star F_7+\frac{2}{3}(\alpha-2\beta)\text{i g}A_3\wedge  F_4\wedge  F_4.
\ee
where $F_4=dA_3$, and $F_7=dA_6-\text{g} A_3\wedge F_4$.
Further physical requirements might constrain these parameters.

The chiral scalar in two dimensions is described by the following pseudo-action and constraint:
\be\label{Schiral0}
S_{\text{chiral}} = \int_{\Sigma_2} \frac{1}{4} F_1 \wedge \star F_1 \,, \quad F_1 = \text{i} \star F_1, \quad dF_1=0.
\ee
Locally it can be written as
\be\label{Schiral}
S_{\text{chiral}} = \int_{\Sigma_2} \frac{1}{4} d\varphi \wedge \star d\varphi \,, \quad d\varphi = \text{i} \star d\varphi,
\ee
where \(\Sigma_2\) is a closed, orientable two-dimensional manifold. This pseudo-action is invariant under a global 0-form symmetry, \(\delta \varphi = \Lambda_0\), with \(d\Lambda_0 = 0\). In the presence of a background \(U(1)\)  field \(A\), we can relax the usual  closure \(d\Lambda_0= 0\), condition  allowing \(d\Lambda_0 \neq 0\). This leads to a  global symmetry given by the combination of transformations \(\delta \varphi = \Lambda_0\) and \(\delta A = d\Lambda_0\). A new action invariant under these transformations is obtained through the  replacement in the action $F_1\rightarrow F_1-A$.
\be
S_0 = \int_{\Sigma_2} \frac{1}{4} (d\varphi - A) \wedge \star (d\varphi - A).
\ee

An additional deformation introduces a non-invariant term into the action, given by
\be
S = S_0 + S^{(d)} = \int_{\Sigma_2} \frac{1}{4} (d\varphi - A) \wedge \star (d\varphi - A) + \text{i} \text{g} A  \wedge d\varphi,
\ee
where  $\text{g}=\frac{1}{2}$. It is worth highlighting one of the most important features of this non-invariant term: the deformation term, $\text{i} \text{g} A  \wedge d\varphi$, is linear in the field \(A\), and more importantly, its transformation does not depend on the scalar field \(\varphi\). Under a variation \(\delta \varphi = \Lambda_0\) , and \( \delta A = d\Lambda_0 \), the change in the action is
\be\label{gauge-trans-S2}
\delta S = \text{i} g \int_{\Sigma_2} A \wedge d\Lambda_0.
\ee
It is important to emphasize that we have applied Stokes’ theorem in the previous integral. This is justified because $A\in\Omega^1(\Sigma_2)$, is globally defined, i.e., we are working on a trivial bundle.

The partition function of the original chiral theory is given by
\be
Z_{\text{chiral}} = \int [D \varphi] \, \text{e}^{-S_{\text{chiral}}},
\ee
and can be obtained from the deformed action by setting \(A = 0\). That is, the partition function with \(A = 0\) is simply
\be
Z[0] = Z_{\text{chiral}},
\ee
and the general partition function with non-zero \(A\) is
\be
Z[A] = \int [D \varphi] \, \text{e}^{-S}.
\ee
Since the deformed action is not invariant under these transformations, the partition function is not a scalar functional but rather a section of a line bundle, whose transformation is given by
\be
Z[A + \delta A] = \text{e}^{-\text{i} \text{g} \int_{\Sigma_2} A \wedge d\Lambda_0} Z[A].
\ee
The  condition of \textit{holomorphy} together with the  \textit{Ward identity}, allow us to establish invariant equations for the partition function. Complex coordinates reveal that the partition function $Z[A]$, is a holomorphic section of a line bundle
\be
S=\text{i}\int_{\Sigma_2}dz\wedge d\bar{z}(\partial_z\varphi\partial_{\bar{z}}\varphi-2\partial_z\varphi A_{\bar{z}}+A_zA_{\bar{z}}),
\ee
\be
\Big(\frac{\delta}{\delta A_z}-\text{i}A_{\bar{z}}\Big)Z=0\label{Holo-Z},
\ee
and
\be
\Big(\partial_{\bar{z}}\big(\frac{\delta}{\delta A_{\bar{z}}}-\text{i}A_{z}\big)+2\text{i}\partial_{z}A_{\bar{z}}\Big)Z=0\label{Ward-Z}.
\ee
The transformation of  $Z$, rewrites as 
\be\label{S-CS gauge transfor}
\delta S=-\text{i}\int_{\Sigma_2}dz\wedge d \bar{z}\big( \partial_z\Lambda_0 A_{\bar{z}}-\partial_{\bar{z}}\Lambda_0 A_z\big)\equiv \Phi[A_z,A_{\bar{z}}],
\ee
\be
Z[A_z + \delta A_z,A_{\bar{z}} + \delta A_{\bar{z}}] = \text{e}^{-\Phi[A_z,A_{\bar{z}}]} Z[A].\label{gauge-complex-coo}
\ee
From \eqref{gauge-complex-coo} we can infer that the space of fields can be considered as the trivial  line bundle   endowed
with the covariant derivatives
\ba\label{covD2d}
\frac{D}{DA_z} & = & \frac{\delta}{\delta A_z}-\text{i}A_{\bar{z}},\quad \frac{D}{DA^{\prime}_z}=\text{e}^{-\Phi}\frac{D}{DA_z},\nonumber\\
\frac{D}{DA_{\bar{z}}} & = & \frac{\delta}{\delta A_{\bar{z}}}+\text{i}  A_{z},\quad \frac{D}{DA^{\prime}_{\bar{z}}}=\text{e}^{-\Phi}\frac{D}{DA_{\bar{z}}}.
\ea
Using the covariant derivatives equations \eqref{Holo-Z} and \eqref{Ward-Z} 
can be written as
\be
\frac{D}{DA_{z}}Z=0.\label{HoloCov}
\ee
and
\be\label{wardD2d}
\Big(\partial_{\bar{z}}\frac{D}{DA_{\bar{z}}}+2\text{i}(\partial_zA_{\bar{z}}-\partial_{\bar{z}}A_z)\Big)Z=0.
\ee

Since previous equations are invariant and
\be \label{curv=0}
\Big[\frac{D}{DA_z(z,\bar{z})},\frac{D}{DA_z(z^{\prime},\bar{z}^{\prime})}\Big]=\Big[\frac{D}{DA_{\bar{z}}(z,\bar{z})},\frac{D}{DA_{\bar{z}}(z^{\prime},\bar{z}^{\prime})}\Big]=0,
\ee
 
\(Z\) is a holomorphic section of a line bundle. Any other holomorphic section of this line bundle can also be regarded as a solution to  equations \eqref{HoloCov} and \eqref{wardD2d}.

Witten's crucial observation is that there exists another section of this line bundle that can be represented as a path integral of the Chern-Simons theory at level $k=\text{g}=\frac{1}{2}$, on a three-dimensional spin manifold \(\Sigma_3\), with boundary \(\Sigma_2\). This section is given by
\be
Z[A] = \int [D a] \, \text{e}^{-\text{i} k \int_{\Sigma_3} a \wedge d a} \,, \quad \partial \Sigma_3 = \Sigma_2 \,, \quad a|_{\partial \Sigma_3} = A,
\ee
and the transformation of the  field is \(\delta a = d\Lambda_0\).   Here $a\in\Omega^1(\Sigma_3)$.  For the sake of simplicity, we have restricted the discussion to trivial bundles and globally defined 1-forms $A$, and $a$. Nevertheless, these results extend to non-trivial bundles and connections; see \cite{Belov:2006jd} and the references therein. We can now obtain \(Z_{\text{chiral}}\) by computing the Chern-Simons wavefunction with vanishing boundary conditions:
\be
Z_{\text{chiral}} = \int [D a] \, \text{e}^{-\text{i} k \int_{\Sigma_3} a \wedge d a} \,, \quad a|_{\partial \Sigma_3} = 0.
\ee

Another obstacle we encounter is when we attempt to couple the  field strengths of M-theory with background fields to make them  invariant under a higher-form  global  symmetry, similarly to the previous case described below \eqref{Schiral}. After identifying a global higher-form symmetry of M-theory on closed orientable 11-dimensional manifolds, we find that while the $F_4$, coupling appears straightforward $F_4\rightarrow F_4-c_4$, \cite{Gaiotto:2014kfa} to find the proper invariant coupling with $F_7$, we need extra assumptions.  One could naively  replace  $F_7\rightarrow F_7-c_7$, the issue with this choice is that in order to get an invariant  object  the transformation of $c_7$, depends on $A_3$. 

As pointed out above equ. \eqref{gauge-trans-S2} for the case of the chiral scalar, in this case if we want to built a 12-dimensional action transforming covariantly under a  transformation involving the background  fields, the transformation of these background  fields should not involve the dynamical fields of the theory. Otherwise we would have to justify its presence in the 12-dimensional holographic bulk. Even worse would be the fact that the partition function built out of   $F_7\rightarrow F_7-c_7$, would not behave as a section of a line bundle. It turns out that the proper transformation for $c_7$, is obtained when we assume a coupling of the form $F_7\rightarrow F_7+2 A_3\wedge c_4-c_7$.\\

The formalism developed in this paper is designed to address the challenge of constructing a quantum theory for democratic M-theory, focusing particularly on the non-chiral form fields and their interactions. While previous approaches have made significant strides in understanding chiral fields in lower dimensions, the extension to higher-dimensional field theories, such as those involved in M-theory, requires a more comprehensive framework that accounts for both electric and magnetic degrees of freedom on an equal footing. In particular, the presence of Chern-Simons terms and  global higher-form symmetries necessitates a refined treatment to ensure consistency in the quantum formalism.

In this work, we propose an extension of the path integral formulation to include the democratic actions with Chern-Simons terms and higher-group symmetries. This extension involves carefully constructing the action in such a way that it remains consistent with the underlying  symmetries while incorporating the necessary duality relations between electric and magnetic field strengths. Furthermore, the formalism is designed to incorporate the necessary physical constraints, ensuring a consistent and robust quantum description of the theory. By utilizing this approach, we not only simplify the calculation of the partition function but also pave the way for future applications to other theories that exhibit similar symmetries and structures,  and the  extension to non-trivial bundles where a more powerful  cohomological treatment is needed \cite{Hopkins:2002rd,Fiorenza:2012mr}.

The paper is organized as follows: in Section 2 we  present the formulation of the democratic M-theory action, including the introduction of Chern-Simons terms and their role in the construction of the quantum theory. Section 3, discusses the coupling of democratic M-theory with background fields and presents the global transformation  of these fields. In Section 4, based on these transformations we find the holographic 12d and  deformed 11d actions. In Section 5 we derive the partition function of the theory and examine its invariance.  We provide enough evidence although not a formal proof that shows that the partition function we are computing is a section of a line bundle. Conclusions, followed by two appendices, are presented in section 6.

\section{Democratic M-theory}

We define as "democratic" those actions that incorporate both electric and magnetic dynamical degrees of freedom (d.o.f.), such that their classical e.o.m. coincide with those of the original theory, where only electric or only magnetic d.o.f. are dynamical, after imposing the duality relation between electric and magnetic field strengths.

We think  that this kind of actions can be defined without making references to the classical theory, hence the e.o.m. Nonetheless, for the scope of this work the definition in previous paragraph is enough.

Our starting point is the Euclidean action:
\be
S_E=\int_{M_{11}}\frac{1}{4}F_4\wedge\star F_4 +\text{i}\frac{\text{g}}{6}A_3\wedge F_4 \wedge F_4,\label{originalMtheory}
\ee
where $F_4=dA_3$, and M-theory corresponds to $\text{g}=-\frac{1}{2}$. Since we are interested in the partition function, $M_{11}$ is considered to be a closed, orientable manifold. The corresponding e.o.m. are given by
\be 
d\star F_4=-\text{i}\text{g}F_4\wedge F_4.
\ee

By defining 
\be
F_7=dA_6-\text{g} A_3\wedge F_4,
\ee
with Bianchi identity 
\be 
dF_7=-gF_4\wedge F_4.
\ee
We see that consistency between the equations of motion for $A_3$, and $A_6$, and their Bianchi identities requires
\be
F_7=-\text{i}\star F_4.\label{dual-cons}
\ee

Based on this, we propose the democratic action
\be\label{demo0}
S_{\text{DEM}}=\int_{M_{11}}\alpha F_4\wedge\star  F_4+\beta F_7\wedge\star F_7+\frac{2}{3}(\alpha-2\beta)\text{i g}A_3\wedge  F_4\wedge  F_4,
\ee
which can be rewritten as 
\be\label{demok}
S_{\text{DEM}}=\int_{M_{11}}\alpha F_4\wedge\star  F_4+\beta F_7\wedge\star F_7+\kappa F_4\wedge F_7,
\ee
where $\kappa=-\frac{2}{3}(\alpha-2\beta)\text{i}$.

\section{Coupling the democratic action with  background fields}

In this section, we treat \eqref{demok} similarly to the self-dual scalar in 2d \cite{Witten:1996hc,Belov:2006jd}, or more precisely, the non-abelian WZ model \cite{Witten:1991mm}. We follow the procedure outlined in \cite{Witten:1996hc,Belov:2006jd,Witten:1991mm}, summarized as follows:
\begin{enumerate}
\item Identify a higher-form global symmetry of \eqref{demok}.
\item Couple with a background field to  relax the  closure condition.
\item Deform the new  invariant action by an appropriate combination of electric and magnetic fields. 
\end{enumerate}

Since we are working on a closed manifold, the M-theory actions \eqref{demo0} and \eqref{demok} enjoy, in addition to the standard gauge symmetry, a higher-form global symmetry given by
\ba\label{gaue-sym-A}
\delta A_3 & = & \Lambda_3,\nonumber\\
\delta A_6 & = & \Lambda_6+\text{g}A_3\wedge \Lambda_3,
\ea
with $d\Lambda_3=d\Lambda_6=0$. Note these transformations leave invariant $F_4$, and $F_7$. For these  fields infinitesimal and large transformation coincide.
This type of global symmetry appeared in the literature by the first time in  \cite{Cremmer:1998px}, but in the context of higher-form global symmetry  was first explored in \cite{Fernandez-Melgarejo:2024ffg}.

Following the arguments given in the introduction, after relaxing the closure  condition  $d\Lambda_3=d\Lambda_6=0$,  we find that a new invariant action can  be written as
\be\label{demo-gauged}
S_{0}=\int_{M_{11}}\alpha {\cal F}_4\wedge\star {\cal F}_4+\beta {\cal F}_7\wedge\star {\cal F}_7+\gamma {\cal F}_4\wedge {\cal F}_7,
\ee
where
\ba
{\cal F}_4 & = & dA_3-c_4,\nonumber\\
{\cal F}_7 & = & dA_6-g A_3\wedge F_4+2\text{g} A_3\wedge c_4-c_7.
\ea
Note that the coefficient in front of the Chern-Simons term is denoted by $\gamma$ rather than $\kappa$,\footnote{This is because the deformed action, unlike the chiral compact scalar case,  also contains a Chern-Simons term.} in section \ref{fullPF} we  fix  this constant. The unusual term $2g A_3\wedge c_4$ in ${\cal F}_7$ is necessary to ensure the correct transformation for $c_7$.  

We briefly remark that our aim is not to modify the structure of M-theory. The background fields 
$c_4$, and $c_7$, are not part of the original theory \eqref{originalMtheory}. Rather, they, together with the 12-dimensional manifold introduced in the next section, serve merely as auxiliary devices to represent the partition function associated with \eqref{demok}.

Infinitesimal transformations of the background  fields read as
\ba\label{gaue-sym-c}
\delta c_4 & = & d\Lambda_3,\nonumber\\
\delta c_7 & = & d\Lambda_6+2\text{g}\Lambda_3\wedge c_4.  
\ea
While large transformations  read as 
\ba\label{gaue-sym-c-large}
 c^{\prime}_4 & = &  c_4+ d\Lambda_3,\nonumber\\
 c^{\prime}_7 & = &c_7+ d\Lambda_6+2\text{g}\Lambda_3\wedge c_4+\text{g}\Lambda_3\wedge d\Lambda_3.  
\ea
As mentioned above,  there is no mixing between the transformations of the $A$ fields and the $c$ fields, which is crucial for defining a well-structured 12-dimensional action based on these  transformations.

With these definitions in hand, we can compute the identities
\ba
\frac{\delta S_0}{\delta c_4} & = &-\Big(2\alpha\star {\cal F}_4+\gamma{\cal F}_7-2g A_3\wedge(2\beta \star{\cal F}_7+\gamma{\cal F}_7)\Big),\nonumber\\
\frac{\delta S_0}{\delta c_7} & = & -(2\beta \star{\cal F}_7+\gamma{\cal F}_7),\nonumber\\
\frac{\delta S_0}{\delta A_6} & = & d(2\beta \star{\cal F}_7+\gamma{\cal F}_7)=-d(\frac{\delta S_0}{\delta c_7}),\nonumber\\
\frac{\delta S_0}{\delta A_3} & = & -d(\frac{\delta S_0}{\delta c_4})-2\text{g}c_4\wedge \frac{\delta S_0}{\delta c_7}+\text{g} A_3\wedge \frac{\delta S_0}{\delta A_6}.
\ea

These relations allow us to derive the Ward identities for $Z[c_4,c_7]$:
\ba
d(\frac{\delta}{\delta c_7})Z[c_4,c_7]   & = & 0,\nonumber\\
\Big(d(\frac{\delta}{\delta c_4})+2\text{g}\,c_4\wedge\frac{\delta}{\delta c_7}\Big) Z[c_4,c_7] & = & 0.
\ea
Previous equations establish the invariance of the partition function  and together with $Z[c^{\prime}_4,c^{\prime}_7]=Z[c_4,c_7]$,  demonstrate that $Z[c_4,c_7]$, behaves as a scalar.

\section{Holographic 12d, and 11d deformed actions }

Before deriving the 11d deformation, we first examine the 12d action and its transformation. We note that the action
\be\label{S12}
S_{12}=\zeta\int_{M_{12}} C_4\wedge d C_7+C_7\wedge dC_4-\frac{2}{3}\text{g}\, C_4\wedge C_4\wedge C_4,
\ee
with $\partial M_{12}=M_{11}$ and
\ba
C_4{|}_{\partial M_{12}} & = & c_4,\nonumber\\
C_7{|}_{\partial M_{12}} & = & c_7,
\ea
transforms linearly in the fields $c_4$ and $c_7$ under
\ba\label{gaue-sym-C}
\delta C_4 & = & d\Lambda_3,\nonumber\\
\delta C_7 & = & d\Lambda_6+2\text{g}\Lambda_3\wedge C_4.  
\ea
Where  \eqref{gaue-sym-C} is the extension to the 12-dimensional bulk of \eqref{gaue-sym-c}.
A similar action was derived in \cite{Evnin:2023ypu}, although through different methods and without using the global symmetry \eqref{gaue-sym-A} and \eqref{gaue-sym-c}. The infinitesimal transformation of the action is given by
\be \label{S12-gauge-transform}
\delta S_{12}=\zeta \int_{M_{11}}c_7\wedge d\Lambda_3-c_4\wedge d\Lambda_6,
\ee
while a large transformation takes the form 
\be \label{S12-gauge-Large-transform}
S^{\prime}_{12}=S_{12}+\Phi[c_4, c_7],
\ee
where
\be
\Phi[c_4, c_7]=\zeta \int_{M_{11}}c_7\wedge d\Lambda_3-c_4\wedge d\Lambda_6+\text{g}\Lambda_3\wedge d\Lambda_3\wedge c_4+\frac{1}{3}\text{g}\Lambda_3\wedge d\Lambda_3\wedge d\Lambda_3.
\ee

Here we would like to highlight the similarity between \eqref{S12-gauge-transform} and \eqref{S-CS gauge transfor}. In fact we could consider \eqref{S12-gauge-transform} as a clue indicating that a similar local holomorphic-antiholomorphic splinting $A=(A_{z}, A_{\bar{z}})$, might exists for a section $c$, of a more general  bundle such that  locally $c=(c_4, c_7)$.

The corresponding 11d deformed action is obtained analogously, guided by \eqref{S12-gauge-transform}.  It is given by
\be
S^{(d)}=\zeta \int_{M_{11}}c_7\wedge F_4-c_4\wedge(dA_6+gA_3\wedge F_4)-\frac{\text{1}}{3}F_4\wedge F_7.
\ee
The infinitesimal transformation of $S^{(d)}$, coincides with \eqref{S12-gauge-transform}.
While the large transformation reads
\be \label{Sd-Large-transform}
S^{\prime (d)}=S^{(d)}+\Phi[c_4, c_7]+\zeta \int_{M_{11}}dX,
\ee
where
\be
X=A_3\wedge  d\Lambda_6-A_6\wedge  d\Lambda_3-\frac{1}{3}\text{g}(\Lambda_3\wedge A_3\wedge d A_3+A_3\wedge \Lambda_3\wedge d\Lambda_3).
\ee
Since $M_{11}$, is closed, last term in  \eqref{Sd-Large-transform}  does not contribute to  the large transformation of $S^{(d)}$.

\section{Full 11D deformed partition function as a section of a line bundle}

The goal of this section is to derive the Ward identities associated with the full 11d deformed action and give some evidence that the partition function might be a section of a line bundle.
\be
S=S_0+S^{(d)}.
\ee 
Since this action is not  invariant, we expect anomalous Ward identities to arise.

We now pose the problem properly. Our objective is to compute the partition function of the democratic M-theory, whose action is given by \eqref{demok}, under the condition \eqref{dual-cons}. However, quantum mechanically the best and sensible we can do is to work under the constraint $\langle F_7+\text{i}\star F_4\rangle=0$.
So,
\be
Z_{DEM}=\int \big[DA_3DA_6 \big]\text{e}^{-S_{DEM}[A_3,A_6]},
\ee
with  the constraint
\be
\langle F_7+\text{i}\star F_4\rangle=\int \big[DA_3DA_6 \big](F_7+\text{i}\star F_4)\text{e}^{-S_{DEM}[A_3,A_6]}=0,\nonumber
\ee
here, instead of computing $Z_{DEM}$ directly, we will use the auxiliary partition  function
\be
Z[c_4,c_7]=\int \big[DA_3DA_6 \big]\text{e}^{-(S_0+S^{(d)})},\nonumber
\ee 
with the condition
\be
Z[0,0]=Z_{DEM}\label{BC}.
\ee
We consider an invariant measure 
\be
\big[DA^{\prime}_3DA^{\prime}_6 \big]=\big[DA_3DA_6 \big].
\ee

A straightforward exercise leads to identities analogous to those derived above, but now for $S$:
\ba
\frac{\delta S}{\delta c_4} & = & \frac{\delta S_0}{\delta c_4} -\zeta(dA_6+gA_3\wedge F_4),\nonumber\\
\frac{\delta S}{\delta c_7} & = & \frac{\delta S_0}{\delta c_7}+\zeta F_4,\nonumber\\
\frac{\delta S}{\delta A_6} & = & \frac{\delta S_0}{\delta A_6}+\zeta\, dc_4,\nonumber\\
\frac{\delta S}{\delta A_3} & = & \frac{\delta S_0}{\delta A_3}+\zeta\left(dc_7-gc_4\wedge F_4+g F_4\wedge F_4-gd(c_4\wedge A_3)\right).
\ea
These can be reorganized into the more compact expressions
\ba
\frac{\delta S}{\delta A_6} & = & d(\frac{\delta S}{\delta c_7})+\zeta\, dc_4,\nonumber\\
\frac{\delta S}{\delta A_3} & = & -d\left(\frac{\delta S}{\delta c_4}\right)-2gc_4 \wedge\frac{\delta S}{\delta c_7}
       +g A_3\wedge\frac{\delta S}{\delta A_6}+\zeta\, dc_7.
\ea

The resulting anomalous Ward identities for the partition function read 
\ba\label{full-ward}
d\left(\frac{\delta}{\delta c_7}-\zeta c_4\right)Z[c_4,c_7]   & = & 0,\nonumber\\
\Big(d\left(\frac{\delta}{\delta c_4}+\zeta c_7\right)+2g\,c_4\wedge\frac{\delta}{\delta c_7}\Big) Z[c_4,c_7] & = & 0.
\ea
According to \eqref{S12-gauge-transform}
\be\label{gaugeT-fullZ}
Z[c_4+\delta c_4,c_7+\delta c_7]=\text{e}^{-\Phi[c_4,c_7]}Z[c_4,c_7],
\ee
under the higher-group symmetry \eqref{gaue-sym-A} and \eqref{gaue-sym-c}.  This transformation together with  these equations show that $Z[c_4,c_7]$ is not a scalar but rather a section of a line bundle. To  present more evidence about this statement we  proceed as in the chiral scalar and identify the covariant derivatives and prove that their corresponding curvatures vanish.

The analysis of the covariant derivatives in this case is a bit more difficult than for the chiral scalar. From \eqref{gaugeT-fullZ} we see that there are two objects that behave as covariant derivatives
\ba\label{covD11d}
\frac{D}{Dc_4} & = & \frac{\delta}{\delta c_4}-\zeta c_7,\nonumber\\
\frac{D}{Dc_7} & = & \frac{\delta}{\delta c_7}+\zeta c_4.
\ea 
We highlight again some similarities with the chiral  scalar, note that \eqref{covD11d} has a similar structure to \eqref{covD2d}. This again indicates that something similar to the holomorphic-antiholomorphic splitting  is happening in this setup.
To prove that these two objects are indeed covariant derivatives we look for their transformation. The expectation is that if these objects are covariant derivatives acting on sections of a line bundle they should transform as the partition function $Z[c_4,c_7]$, similar to the chiral scalar \eqref{covD2d}.

A quick inspection however reveals that under an infinitesimal transformation \eqref{gaue-sym-c} the covariant derivatives (acting on $Z$) transform as
\ba\label{D-transf}
\frac{D}{Dc_4^{\prime}} & = & \text{e}^{-\Phi}\Big(\frac{D}{Dc_4}-2\text{g}\Lambda_3\wedge \frac{D}{Dc_7} \Big),\nonumber\\
\frac{D}{Dc_7^{\prime}} & = & \text{e}^{-\Phi}\frac{D}{Dc_7},
\ea 
where
\ba\label{gauge-prime}
c^{\prime}_4 & = & c_4+ d\Lambda_3,\nonumber\\
c^{\prime}_7 & = & c_7 +d\Lambda_6+2\text{g}\Lambda_3\wedge c_4.  
\ea
The first transformation may be seen as a drawback when compare with the expectation. Fortunately it is not. To properly understand this transformation we have to take into account that transformation \eqref{gauge-prime} mixes $c_4$, and $c_7$. So, what we are looking at in \eqref{D-transf} is the correct covariant  transformation of the covariant derivative. We have the proper expected phase $\text{e}^{-\Phi}$, and the extra term in \eqref{D-transf}. Put simply, it is just a consequence of the chain rule for the derivative, see Apendix A.

Now we can compute their corresponding curvatures, showing that as expected for the identification of the line bundle, they vanish
\be 
\Big[\frac{D}{Dc_7(x)},\frac{D}{Dc_7(x^{\prime})}\Big]=\Big[\frac{D}{Dc_4(x)},\frac{D}{Dc_4(x^{\prime})}\Big]=0.
\ee

Using the covariant derivatives, equation \eqref{full-ward} can be written in a more manifest, see Appendix A, invariant form
\be \label{WardD4}
\Big(d\frac{D}{Dc_7}-2\zeta G_5 \Big)Z=0,
\ee
and
\be\label{WardD7}
\Big(d\frac{D}{Dc_4}+2\text{g}c_4\wedge\frac{D}{Dc_7}+2\zeta G_8\Big)Z=0,
\ee
where
\ba
 G_5  & = & d c_4,\nonumber \\
 G_8 & = & dc_7-\text{g}c_4 \wedge c_4,
\ea
are the fields strength associated to the c-fields.
Thus,  $Z[c_4,c_7]$, could be a section of a line bundle over the space of all connections $(c_4,c_7)$.

Although we believe we have made significant progress in understanding the global structure of the  $A_3+A_6$, system and more  important the  $C_4+C_7$, system, see section \ref{sec-heisengerg}, we consider that  we have not  formally proven that such a particular line bundle exists. This will be proven elsewhere.

Nonetheless, we have enough evidence to move forward even lacking this formal prove. Rigorously proving that such a line bundle exists and  $Z[c_4,c_7]$, is indeed a section of  this  line bundle allows equations \eqref{WardD4} and \eqref{WardD7} to be solved straightforwardly. A solution (by construction) is given by
\be\label{Holo-parti-Z}
Z[c_4,c_7]=\int [DC_4\,DC_7] \exp\Big(\zeta\int_{M_{12}} C_4\wedge dC_7+C_7\wedge dC_4-\frac{2}{3}g\, C_4\wedge C_4\wedge C_4\Big),
\ee
with boundary conditions
\ba
C_4{|}_{\partial M_{12}} & = & c_4,\nonumber\\
C_7{|}_{\partial M_{12}} & = & c_7.
\ea 

\subsection{Full partition function and the constraint  $F_7=-\text{i}\star F_4$\label{fullPF}}

So far, we have not fully determined the partition function we are interested in. The constants $\gamma$ and $\zeta$ remain undetermined. Recall that we are working under the condition \eqref{BC}. By setting $c_4=c_7=0$, and comparing  the action $S=S_0+S^{(d)}$, written in terms of $\gamma$,  see \eqref{demo-gauged}, with  $S_{DEM}$ \eqref{demok}, which is written in terms  of  $\kappa$, we find that these constants  should be related as
\be
\gamma-\frac{1}{3}\zeta=\kappa=-\frac{2}{3}(\alpha-2\beta)\text{i}\label{sys1}.
\ee
This condition ensures that we are effectively  computing the  partition  function associated to $S_{DEM}$, trough the auxiliary action $S=S_0+S^{(d)}$.

On the other hand, there is an integrability condition for the Ward identity \eqref{full-ward}. Taking the derivative of the second equation and using the first one in \eqref{full-ward}, we find (assuming $dc_4\neq 0$)
\be
\left(\frac{\delta}{\delta c_7}+\zeta c_4\right)Z[c_4,c_7]=0.
\ee
Which can  be suggestively  written as  
\be\label{Holocov-C}
\frac{D}{Dc_7}Z[c_4,c_7]=0.
\ee
This equation is the analogous of equation \eqref{HoloCov} for the chiral compact scalar.

To clarify the meaning of this equation, we compute it explicitly:
\be
\int [DA_3 DA_6]\left(\star {\cal F}_7+\frac{\gamma-\zeta}{2\beta}{\cal F}_4\right)\text{e}^{-S}=0.
\ee
At this stage, it is straightforward to see that if we set 
\be
\frac{\gamma-\zeta}{2\beta}=\text{i}\label{sys2},
\ee
we obtain 
\be
\langle \star {\cal F}_7+\text{i}{\cal F}_4 \rangle=0.
\ee
What is remarkable about this equation is that when $c_4=c_7=0$, it reduces to the constraint
\be
\langle \star  F_7+\text{i}F_4 \rangle=0,
\ee 
which is the quantum version of the classical constraint $\star F_7=-\text{i}F_4$. With \eqref{sys1} and \eqref{sys2}, we can fully fix the constants and, therefore, the partition function:
\ba
\gamma & = & -(\alpha-\beta)\text{i},\nonumber 	\\
\zeta & = & -(\alpha+\beta)\text{i}.
\ea
It is worth recalling that the only input we assume is $\alpha$ and $\beta$, and after fixing them, we can fully determine the democratic action and the partition function.

\subsection{Global structure  of  $A_3+A_6$, and $C_4+C_7$, systems and the Heisenberg group\label{sec-heisengerg}}

In this section we study the algebra  and the  corresponding group   defined by the transformations of the A, and C fields.
For the A, fields we have
\ba
\delta A_3 & = & \Lambda_3,\nonumber\\
\delta A_6 & = & \Lambda_6+\text{g}A_3\wedge \Lambda_3,
\ea
where infinitesimal and large transformations coincide. For the C,  fields  the infinitesimal transformation is given by
\ba
\delta C_4 & = & d\Lambda_3,\nonumber\\
\delta C_7 & = & d\Lambda_6+2\text{g}\Lambda_3\wedge C_4,
\ea
while the large transformation reads
\ba\label{C-LARGE}
 C^{\prime}_4 & = & C_4+ d\Lambda_3,\nonumber\\
 C^{\prime}_7 & = & C_7+ d\Lambda_6+2\text{g}\Lambda_3\wedge C_4+\text{g}\Lambda_3\wedge d\Lambda_3.  
\ea

After composing the transformations $1\to2$ and $2\to3$ (equivalently $1\to3$), we obtain the following. 
In particular, the $2\to3$ transformation reads
\ba
A_3^{(3)} & = &  A_3^{(2)} +\Lambda_3^{(23)},\nonumber\\
A_6^{(3)} & = & A_6^{(2)}+\Lambda_6^{(23)}+\text{g}A_3^{(2)}\wedge \Lambda_3^{(23)}.
\ea
The transitivity of the composition implies
\ba\label{OJO}
\Lambda_3^{(13)}  & = &   \Lambda_3^{(12)}+\Lambda_3^{(23)},\nonumber\\
\Lambda_6^{(13)}  & = &   \Lambda_6^{(12)}+\Lambda_6^{(23)}+\text{g}\Lambda_3^{(12)}\wedge \Lambda_3^{(23)}.
\ea
Therefore, the system $A_3+A_6$ is globally well defined provided that the parameters satisfy \eqref{OJO}. 
Repeating the same analysis for the large transformations of the $C$ fields \eqref{C-LARGE}, one finds that the $C_4+C_7$ system is likewise globally well defined, provided that \eqref{OJO} holds.

Composition laws \eqref{OJO} resemble those of a Heisenberg group. To expose the group structure underlying \eqref{OJO} we define the set
\be
\text{G}_{\text{H}}=\Omega^3(M_{11})\oplus\Omega^6(M_{11}).
\ee
An element of this set is of the form $(\Lambda_3,\Lambda_6)$, and given two such elements $(\Lambda_3,\Lambda_6)$, and $(\Sigma_3,\Sigma_6)$,  the product is defined as 
\be
(\Lambda_3,\Lambda_6)\star(\Sigma_3,\Sigma_6)=(\Lambda_3+\Sigma_3,\Lambda_6+\Sigma_6+\text{g}\Lambda_3\wedge\Sigma_3)\in \text{G}_{\text{H}}.
\ee
The identity element is 
\be
\mathbb{I}=(0,0),
\ee
while the inverse element is given by
\be
(\Lambda_3,\Lambda_6)^{-1}=(-\Lambda_3,-\Lambda_6).
\ee
Having identified the group structure of the large transformations, the algebra follows immediately. Now we identify the algebra as the  set   
\be
\text{g}_{\text{H}}=\Omega^3(M_{11})\oplus\Omega^6(M_{11}).
\ee
The Lie bracket is given by
\be
\big[(\Lambda_3,\Lambda_6),(\Sigma_3,\Sigma_6)\big]=(0,2\text{g}\Lambda_3\wedge\Sigma_3).
\ee
This completes the proof that the systems $A_3+A_6$, and $C_4+C_7$, together with their transformation laws, define globally well-defined objects with an underlying Heisenberg-type group structure.

\section{Conclusions}

In this paper, we have presented a formal and quantum treatment of the democratic formulation of M-theory, focusing particularly on the holographic partition function. We have explored the complexities arising from non-chiral form fields and the incorporation of Chern-Simons terms and higher-group  symmetries within the M-theory framework. By addressing these challenges, we have contributed to the development of a more complete and consistent understanding of democratic actions in higher-dimensional theories, emphasizing their quantum mechanical foundations.

One of the primary contributions of this work is the clarification of the relationship between the two predominant approaches to formulating democratic M-theory: the formal quantum treatment, and the more classical, edge-mode description. These two approaches offer complementary insights into the nature of the gauge fields and their interactions with the bulk manifold. While the first approach, based on path integrals over topological Chern-Simons theories, provides an elegant holographic interpretation, it does not directly address the physical significance of the bulk. The second, more classical approach, which describes chiral form fields and democratic non-chiral forms as edge modes, offers a more physically intuitive picture, but it remains challenging to reconcile with quantum mechanics in its simplest form.

Through the quantum treatment presented in this work, we have managed to bridge this gap by establishing a formalism that incorporates both classical and quantum perspectives. By introducing non-chiral actions with Chern-Simons terms and higher-group symmetries, we extend the established framework for the description of M-theory, providing a more holistic understanding that encompasses both electric and magnetic degrees of freedom. This is a critical step forward, as it allows us to work with a consistent and unified description of the theory, avoiding the pitfalls of non-polynomial or gauge-fixed actions and introducing a robust foundation for further quantum computations.

In particular, the partition function of the democratic M-theory, as derived in this paper, provides a key insight into the nature of the theory's quantum structure. By computing the partition function in the context of the democratic formulation, we obtain not only a tool for understanding the dynamics of the system but also a concrete result that can be used to test and further refine the theoretical framework. The analysis of the field strengths and their invariant couplings is an important step in this process, ensuring that the partition function behaves consistently under global transformations. This consistency is critical, as it guarantees that the partition function corresponds to a valid physical quantity, transforming as a section of a line bundle, rather than as a scalar functional.

Another key result of this work is the introduction of the democratic M-theory action, which incorporates both electric and magnetic degrees of freedom in a manner that is consistent with the underlying  symmetries. This action, which is parametrized by the parameters $\alpha$ and $\beta$, represents a generalization of previous formulations and provides a more flexible and comprehensive framework for understanding the interactions of higher-form fields within M-theory. By working with this more general action, we are able to address the duality relations between electric and magnetic fields, ultimately deriving a quantum partition function that adheres to the necessary constraints.

The results presented here also have broader implications for the study of higher-dimensional gauge theories. The framework we have developed for M-theory is not limited to this specific case but can be applied to other theories with similar structures, such as democratic Maxwell-Chern-Simons theory in 5d, or type IIB $SL(2)$ democratic supergravity. This opens up new avenues for research in the study of quantum field theories and supergravity, particularly in the context of democratic formulations that unify the electric and magnetic degrees of freedom.

Moreover, this paper highlights the importance of the global symmetry  transformations and the covariant derivatives that act on the sections of the line bundle. The introduction of these tools is essential for understanding the transformation properties of the partition function, providing a rigorous mathematical framework for analyzing its behavior under these kind transformations. The analysis of the curvatures associated with these covariant derivatives further establishes the validity of the proposed formulation, ensuring that the partition function remains well-defined and invariant.

There are open questions that need to be addressed, some of them are determinant to settle the validity of these results. Action \eqref{Holo-parti-Z} needs special attention. Although we have several evidences that show this integral is well defined on  a  closed 12d manifold (modulo $2\pi\mathbb{Z}$). Namely, independent of the choice of the manifold,  similar to the Chern-Simons action in 3d,  we have not formally proven this yet.

If ultimately this approach proves to be correct the next natural step would be to introduce the charged objects of the theory and compute the expectation value of the corresponding observable. Although for this we will  need a more powerful cohomological approach similar to \cite{Witten:1996md,Hopkins:2002rd,Fiorenza:2012mr}.

It is straightforward to write down  the gauge invariant object related to the $M_2$ brane in the presence of the higher gauge symmetry and background  fields, it is \cite{Gaiotto:2014kfa}
\be\label{WC4}
W_2[A_3]=\text{exp}\Big[\text{i}\int_{\Sigma_3}A_3-\text{i}\int_{\Sigma_4}c_4\Big],\quad \partial \Sigma_4=\Sigma_3.
\ee
For the $M_5$ brane things are more involve, we believe that the proper object has the form
\be\label{WC7}
W_5[A_3,A_6]=\int [D {\cal H}_3]\text{exp}\Big[\text{i}\int_{\Sigma_6}A_6+\text{i}\int_{\Sigma_7}\big(-\text{g}{\cal H}_3\wedge d{\cal H}_3+2\text{g}{\cal H}_3\wedge c_4-c_7\big) \Big],\quad \delta {\cal H}|_{\partial \Sigma_7}=A_3,
\ee
where $\partial \Sigma_7=\Sigma_6$. Here  we first have used  the Witten representation \cite{Witten:1996hc} of the effective partition function for the $M_5$, brane and then we have couple it with the backgrounds fields. This object is gauge invariant provided $\delta {\cal H}=\Lambda_3$. Note that  the background fields appear in \eqref{WC4} and \eqref{WC7} in a similar way they appear on the field strengths ${\cal F}_4$, and ${\cal F}_7$. An analysis of the consequences of introducing these objects is left for a future work.

In conclusion, the work presented in this paper represents a significant step forward in our understanding of the democratic formulation of M-theory, providing a rigorous and comprehensive framework for the quantum treatment of higher-form fields and their interactions. By addressing the challenges of duality, global higher-form symmetry  invariance, and the coupling of background fields, we have laid the groundwork for future research in this area, opening up new possibilities for exploring the quantum structure of M-theory and other higher-dimensional gauge theories.

\section*{Acknowledgments}
We are grateful to  Iñaki~Garc{\'\i}a Etxebarria and Federico Bonetti for several discussions, comments and suggestions.

\appendix

\section{Invariance of the Ward identities}

In this appendix we show the  invariance of the Ward identities \eqref{full-ward}. To do it systematically, we fist derive how the covariant derivatives \eqref{covD11d}   transform \eqref{D-transf}. Them using these transformation laws we will prove the invariance of the equations \eqref{WardD4} and \eqref{WardD7}, which are written in terms of the covariant derivatives.

We start with the partition function written in the $(c_4^{\prime},c_7^{\prime})$, coordinates,  where they are related with the coordinates $(c_4,c_7)$, as in \eqref{gauge-prime},
\be
Z[c_4^{\prime},c_7^{\prime}]=\int \big[DA^{\prime}_3DA^{\prime}_6 \big]\text{e}^{-S[c_4^{\prime},c_7^{\prime}]},
\ee
where
\be
S[c_4^{\prime},c_7^{\prime}]=S[c_4,c_7]+\Phi[c_4,c_7],
\ee
and 
\be
\big[DA^{\prime}_3DA^{\prime}_6 \big]=\big[DA_3DA_6 \big].
\ee
Computing the derivative in the prime coordinates in term of the non-prime coordinates of the partition function yields to 
\be 
\frac{\delta}{\delta c_4^{\prime}}Z[c_4^{\prime},c_7^{\prime}]=-\int \big[DA_3DA_6 \big] \frac{\delta}{\delta c_4^{\prime}}\Big(S[c_4,c_7]+\Phi[c_4,c_7]\Big)\text{e}^{-(S[c_4,c_7]+\Phi[c_4,c_7])}
\ee
Using the chain rule for the derivative we arrive at 
\be 
\frac{\delta}{\delta c_4^{\prime}(x)}Z[c_4^{\prime},c_7^{\prime}]=-\int \big[DA_3DA_6 \big]\Big( \frac{\delta}{\delta c_4(x)}\big(S+\Phi\big)+\int_{x^{\prime}}\frac{\delta}{\delta c_7(x^{\prime})}\big(S+\Phi\big)\wedge\frac{\delta c_7(x^{\prime})}{\delta c_4^{\prime}(x)}\Big)\text{e}^{-(S+\Phi)}.
\ee
From the inverse transformation of \eqref{gauge-prime} we find
\be
\frac{\delta c_7(x^{\prime})}{\delta c_4^{\prime}(x)}=-2\text{g}\Lambda_3\delta^{11}(x-x^{\prime}).
\ee
Finally
\be\label{D4Z}
\frac{\delta}{\delta c_4^{\prime}}Z[c_4^{\prime},c_7^{\prime}]=\text{e}^{-\Phi}\Big(\big( \frac{\delta Z}{\delta c_4}+\zeta d\Lambda_6 Z \big)-2\text{g}\Lambda_3\wedge  \frac{\delta Z}{\delta c_7} \Big).
\ee
We can proceed similarly for the other derivative to get
\be\label{D7Z}
\frac{\delta}{\delta c_7^{\prime}}Z[c_4^{\prime},c_7^{\prime}]=\text{e}^{-\Phi}\Big(\frac{\delta Z}{\delta c_7}-\zeta d\Lambda_3 Z  \Big).
\ee

From \eqref{D7Z} we can straightforwardly  show that the derivative $\frac{D}{Dc_7} =  \frac{\delta}{\delta c_7}+\zeta c_4$, transform as
\be\label{D4trannsapp}
\frac{D}{Dc_7^{\prime}}Z[c_4^{\prime},c_7^{\prime}]  = \text{e}^{-\Phi}\frac{D}{Dc_7}Z[c_4,c_7].
\ee 
For the other covariant derivative $\frac{D}{Dc_4} =  \frac{\delta}{\delta c_4}-\zeta c_7$, using  \eqref{D4Z}, we can easily show  that  
\be\label{D7trannsapp}
\frac{D}{Dc_4^{\prime}}Z[c_4^{\prime},c_7^{\prime}]  =  \text{e}^{-\Phi}\Big(\frac{D}{Dc_4}-2\text{g}\Lambda_3\wedge \frac{D}{Dc_7} \Big)Z[c_4,c_7].
\ee

Now we are in a condition of showing the  invariance of the Ward identities \eqref{WardD4} and  \eqref{WardD7}. The  invariance of \eqref{WardD4} follows immediately from \eqref{D4trannsapp} and $\delta G_5=\delta (d c_4)=0$. The invariance of \eqref{WardD7} is a bit more involve but after  taking into account \eqref{D7trannsapp} and $\delta G_8=\delta ( dc_7-\text{g}c_4\wedge c_4)=-2\text{g}\Lambda_3\wedge G_5$, we can show that indeed equation \eqref{WardD7} is invariant.

\section{$S_{12}$, as a topological and  well defined action}

In this appendix we show how to properly define \eqref{S12}. We follow similar strategy as in the Chern-Simons theory. Here $M_{12}$, is regarded as a closed manifold and  as the boundary of a 13-dimensional manifold $M_{13}$. Now we set
\be\label{13-int}
I[M_{13}]=\int_{M_{13}}2G_5\wedge G_8,
\ee
where $M_{13}$, is a 13-dimensional close manifold. Note that under the integral sign in previous integral we have what locally would  be the derivative of the 12-dimensional Lagrangian  i.e.,
\be
2G_5\wedge G_8=d\big(C_4\wedge d C_7+C_7\wedge dC_4-\frac{2}{3}\text{g}\, C_4\wedge C_4\wedge C_4\big)
\ee

To check whether $S_{12}$, is
well defined i.e., independent of the choice of $M_{13}$, we consider
two different extensions, involving 13-manifolds $N_{13}$ and $N^{\prime}_{13}$, each with boundary $M_{12}$, and   we  need to show that $I[N_{13}]$, and $I[N^{\prime}_{13}]$
are equal modulo $2\pi$. If $M_{13}$,  is the closed 13-manifold obtained by gluing $N_{13}$ and $N^{\prime}_{13}$, along their common boundary $M_{12}$, (with opposite orientations) we have the relation
\be\label{well-defined}
I[N^{\prime}_{13}]-I[N_{13}]=2\int_{M_{13}}G_5\wedge G_8,
\ee
where $dG_5=0$, and $dG_8+2\text{g}C_4\wedge G_5=0$, such that locally
\ba
G_5 & = & dC_4,\nonumber\\
G_8 &  = & dC_7-\text{g}C_4\wedge C_4.
\ea

Now we need to study integral \eqref{13-int}. Using
\be
\delta G_8=-2\text{g}\Lambda_3\wedge G_5,
\ee

it is easy to check that this integral is gauge invariant, which ultimately will be a consequence of $G_5\wedge G_5=0$. It is also straightforward to show, using only  the Bianchi identities  for $G_5$, and $G_8$,  that $G_5\wedge G_8$, is closed,  $d(G_5\wedge G_8)=0$. However we have not been able to prove that the integral is quantized. We leave  this exercise  for a future work.

\end{document}